\documentclass[11pt,a4paper]{article}
\usepackage{amsfonts,amssymb,mathrsfs}
\usepackage{amsopn}
\usepackage{amsmath,amsthm}
\usepackage{verbatim}

\DeclareMathAlphabet{\mathpzc}{OT1}{pzc}{m}{it}

\theoremstyle{plain}

\theoremstyle{remark}

\title{Fundamental Structures\\of M(brane) Theory}

\author{Jens Hoppe}

\date{\scriptsize{Department of Mathematics, Royal Institute of Technology\\100 44 \ Stockholm, Sweden}}

\begin{document}

\maketitle

\begin{abstract}
	A dynamical symmetry, as well as special diffeomorphism
	algebras generalizing the Witt-Virasoro algebra, 
	related to Poincar\'e-invariance and crucial with regard to
	quantisation, questions of integrability, and M(atrix) theory,
	are found to exist in the theory of relativistic extended
	objects of any dimension.
\end{abstract}

\vspace{1.5cm}

\setlength\arraycolsep{2pt}
\def\arraystretch{1.2}

\noindent
The simplicity of classical string theory, 
and decades of presenting it from one and the same point of view,
have made it difficult to realize some of the central features of
relativistic extended objects 
(described below in the light cone gauge), namely:
\begin{itemize}
\item 
	relativistic invariance implying the existence of a dynamical
	symmetry (irrespective of the dimension of the extended object),
	and
\item 
	the Virasoro algebra being just the simplest example of
	certain (extended) infinite-dimensional diffeomorphism algebras reappearing,
	after gauge fixing (and on the constrained phase space), in the reconstruction of $x^-$.
\end{itemize}

For the purpose of this note, I will restrict myself to the purely
bosonic theory \cite{Hoppe,Goldstone}, i.e.
(analogous results for the supersymmetrized theory will easily follow)
\begin{eqnarray}
	H = \frac{1}{2\eta} \int_{\Sigma_0} \frac{\vec{p}^2 + g}{\rho} \thinspace d^M\varphi
	= H[\vec{x},\vec{p};\eta,\zeta_0] = \int \mathcal{H} \thinspace d^M\varphi, \\
	\qquad g = \det \left( 
		\frac{\partial \vec{x}}{\partial \varphi^a} 
		\cdot \frac{\partial \vec{x}}{\partial \varphi^b} 
		\right)_{a,b=1,\ldots,M}, \nonumber
\end{eqnarray}
with $\rho(\varphi)$ being a non-dynamical density of weight one
(i.e. $\int_{\Sigma_0} \rho(\varphi) \thinspace d^M\varphi = 1$),
$x_i$ and $p_j$ ($i,j=1,\ldots,d=D-2$) canonically conjugate fields
satisfying 
\begin{equation}  \label{divergencefree}
        \int f^a \thinspace \vec{p} \cdot \partial_a \vec{x} \thinspace d^M\varphi = 0 \quad \textrm{whenever} \quad \nabla_a f^a = 0
\end{equation}
for the consistency of

\begin{equation} \label{int_condition_space}
	\eta \partial_a \zeta = \frac{\vec{p}}{\rho} \cdot \partial_a \vec{x}
\end{equation}
which, together with
\begin{equation} \label{int_condition_time}
	2\eta^2 \dot{\zeta} = \frac{{\vec{p} \thinspace}^2 + g}{\rho^2}
\end{equation}
(that actually can also be thought of as defining $\eta$ and $\rho$
in terms of the initial parametrized shape, and the velocity, of the
time-dependent $M$-dimensional extended object moving in $D$-dimensional
Minkowski space),
determine $\zeta$ (usually called $x^-$) up to 
$\zeta_0 = \int \zeta \rho \thinspace d^M\varphi$; 
the time independent positive degree of freedom $\eta$ 
(usually called $P^+$) is canonically conjugate to $-\zeta_0$.

In the mid-eighties, Goldstone
(when proving that the above description is fully Poincar\'e-invariant
\cite{Goldstone}) solved  \eqref{int_condition_space} (assuming
\eqref{divergencefree}) in the form
\begin{equation} \label{zeta_solved}
	\zeta(\varphi) = \zeta_0 + \frac{1}{\eta}
	\int G(\varphi,\tilde{\varphi}) \tilde{\nabla}^a 
	\left( \frac{\vec{p}}{\rho} \cdot \tilde{\nabla}_a \vec{x} \right)\!(\tilde{\varphi})
	\thinspace \rho(\tilde{\varphi}) \thinspace d^M\tilde{\varphi}
\end{equation}
with ($\nabla_a$ being the covariant derivative, and $\Delta$ the Laplacian on $\Sigma_0$)
\begin{equation} \label{Greens_function}
	\int G(\varphi,\tilde{\varphi}) \rho(\varphi) \thinspace d^M\varphi = 0,
	\qquad
	\Delta_{\tilde{\varphi}} G(\varphi,\tilde{\varphi}) = \frac{\delta(\varphi,\tilde{\varphi})}{\rho(\varphi)} - 1.
\end{equation}
Later it will turn out to be useful to slightly
(though, with regard to a variety of aspects: crucially)
rewrite \eqref{zeta_solved} as
\begin{eqnarray} \label{zeta_solved_Laplace}
	\zeta(\varphi) = \zeta_0 + \frac{1}{2\eta} \vec{p}\cdot\vec{x} 
	+ \frac{1}{2} \int G(\varphi,\tilde{\varphi})
	\left( \frac{\vec{p}}{\rho} \cdot \Delta \vec{x} - \vec{x} \cdot \Delta \frac{\vec{p}}{\rho} \right)\!(\tilde{\varphi}) 
	\thinspace \rho(\tilde{\varphi}) \thinspace d^M\tilde{\varphi}
\end{eqnarray}
(splitting $\zeta-\zeta_0$ into parts symmetric resp. antisymmetric
with regard to interchanging $\vec{x}$ and $\vec{p}$, and involving only
the invariant Laplace operator).
I can now present the two key features that I recently found:

\subsection*{Dynamical Symmetry}

When separating the zero-modes
\begin{equation}
	\zeta_0, \qquad \eta, 
	\qquad X_i = \int x_i \rho \thinspace d^M\varphi,
	\qquad P_i = \int p_i \thinspace d^M\varphi
\end{equation}
from the internal degrees of freedom,
\begin{equation}
	x_{i\alpha} := \int Y_\alpha(\varphi) x_i(\varphi) \rho(\varphi) \thinspace d^M\varphi,
	\qquad
	p_{i\alpha} := \int Y_\alpha(\varphi) p_i(\varphi) \thinspace d^M\varphi,
\end{equation}
--- letting $\{ Y_{\alpha} \}_{\alpha=1}^\infty$ be a (together with $Y_{0}=1$)
complete orthonormal set of eigenfunctions on $\Sigma_0$
(conveniently chosen as eigenfunctions of $\Delta$),
\begin{equation} \label{ON-basis}
	\int Y_\alpha Y_\beta \rho \thinspace d^M\varphi = \delta_{\alpha\beta},
	\ \ \sum_{\alpha=1}^\infty Y_\alpha(\varphi) Y_\alpha(\tilde{\varphi}) 
	= \frac{\delta(\varphi,\tilde{\varphi})}{\rho(\varphi)} - 1,
	\ \ \Delta Y_\alpha = -\mu_\alpha Y_\alpha
\end{equation}
--- the Lorentz-invariance of the theory,
in particular implying that
\begin{equation}
	M_{i-} := \int (x_i \mathcal{H} - \zeta p_i) \thinspace d^M\varphi
\end{equation}
satisfies
\begin{equation} \label{Lorentz_minus}
	\{ M_{i-}, M_{j-} \} = 0,
\end{equation}
necessitates that the purely internal contributions
\begin{eqnarray} \label{def_Mb_minus}
	2\eta \mathbb{M}_{i-} := \int (x_i \tilde{\mathcal{H}} - \tilde{\zeta} p_i) \thinspace d^M\varphi
	= x_{i\alpha}\tilde{\mathcal{H}_\alpha} - \tilde{\zeta}_\alpha p_{i\alpha},
\end{eqnarray}
with
\begin{eqnarray} \label{zetatilde}
	\tilde{\mathcal{H}}_\alpha := \vec{p}_\beta \cdot \vec{p}_\gamma 
	\int Y_\alpha Y_\beta Y_\gamma \rho \thinspace d^M\varphi
	+ \int Y_\alpha \frac{g}{\rho} \thinspace d^M\varphi
	=: \vec{p}_\beta \cdot \vec{p}_\gamma d_{\alpha\beta\gamma} + W_\alpha,
\end{eqnarray}
$$
	\tilde{\zeta}_\alpha := 2\eta(\zeta_\alpha - \vec{P} \cdot \vec{x}_\alpha),
	\qquad
	\zeta_\alpha := \int Y_\alpha \zeta \rho \thinspace d^M\varphi,
	\nonumber
$$
satisfy
\begin{equation} \label{dynamical_symmetry}
	\{ \eta\mathbb{M}_{i-}, \eta\mathbb{M}_{j-} \} = \mathbb{M}^2 \thinspace \mathbb{M}_{ij},
	\qquad i,j = 1,\ldots,d,
\end{equation}
where
\begin{equation}
	\mathbb{M}_{ij} := x_{i\alpha} p_{j\alpha} - x_{j\alpha} p_{i\alpha}
\end{equation}
are the generators of internal transverse rotations, and
\begin{equation} \label{mass_squared}
	\mathbb{M}^2 = \tilde{\mathcal{H}}_\alpha \tilde{\mathcal{H}}_\alpha 
	= 2\eta H - \vec{P}^2
\end{equation}
is the square of the relativistically invariant `internal mass',
commuting with $M_{i-}$, $M_{ij}$, $H$, $\vec{P}$, and $\eta$, 
as well as $\eta \zeta_0$ and $\eta X_i$.
\eqref{dynamical_symmetry} is  a simple (but crucial) consequence of
\eqref{Lorentz_minus}, as the parts of $M_{i-}$ that do involve the
zero-modes satisfy
\begin{equation}
	\left\{ X_i H - \zeta_0 P_i + \frac{\mathbb{M}_{ik}}{\eta}P_k \ , \ 
	X_j H - \zeta_0 P_j + \frac{\mathbb{M}_{jl}}{\eta}P_l \right\} 
	= - \frac{\mathbb{M}^2}{\eta^2} \mathbb{M}_{ij},
\end{equation}
which easily follows from $\{H,\mathbb{M}_{ik}\} = 0$,
$\{\zeta_0,\eta\} = -1$, $\{X_i,P_j\} = \delta_{ij}$, and
\begin{equation} \label{Lorentz_spatial}
	\{ \mathbb{M}_{ik}, \mathbb{M}_{jl} \} = -\delta_{kj} \mathbb{M}_{il} \pm \textrm{3 more},
\end{equation}
and \eqref{mass_squared}.
Finally, one checks that
\begin{equation} \label{Lorentz_mixed}
	\{ \eta \mathbb{M}_{i-}, \mathbb{M}_{ij} \} = -\eta \mathbb{M}_{j-},
\end{equation}
and that $\mathbb{M}^2$ commutes with $\eta\mathbb{M}_{i-}$
(and $\mathbb{M}_{ij}$).

This sign of integrability / dynamical symmetry
($\mathbb{M}^2$ appearing in the structure constants of a symmetry
algebra of itself) should be extremely useful for the further 
understanding of relativistic extended objects. 
E.g. if it is possible to promote \eqref{dynamical_symmetry},
\eqref{Lorentz_spatial}, \eqref{Lorentz_mixed}
to commutation relations for corresponding quantum operators
(commuting with $\hat{\mathbb{M}}^2$), one may be able to
calculate the spectrum of $\hat{\mathbb{M}}^2$
purely algebraically
in terms of the Casimirs of the algebra spanned by 
$L_{ij} := \mathbb{M}_{ij}$ and $L_{i,d+1} := \frac{\eta\mathbb{M}_{i-}}{\sqrt{\mathbb{M}^2}}$,
just as in the case of the $d$-dimensional Hydrogen atom,
which is actually \emph{very} close to the relations that I just
derived; the difference lying in the explicit a-priori
relations between the angular momentum $\mathbb{M}_{ij}$
and the generalized Laplace-Runge-Lenz vector
(most likely these relations exist here as well, encoding the dimensionality
and topology of the extended object).
One way to find them is to understand the interplay of the different
diffeomorphism subalgebras involving the totally symmetric structure constants
$d_{\alpha\beta\gamma}$ (cp. \eqref{zetatilde}),
\begin{equation} \label{e}
	e_{\alpha\beta\gamma} := \frac{\mu_\beta - \mu_\gamma}{\mu_\alpha} d_{\alpha\beta\gamma}
\end{equation}
and
\begin{equation}
	g_{\alpha\alpha_1 \ldots \alpha_M} 
	:= \int_{\Sigma_0} Y_\alpha \epsilon^{a_1 \ldots a_M} 
	\frac{\partial Y_{\alpha_1}}{\partial \varphi^{a_1}} \ldots \frac{\partial Y_{\alpha_M}}{\partial \varphi^{a_M}}
	\thinspace d^M\varphi,
\end{equation}
--- part of which I will now come to.

\subsection*{$\mathscr{L}$-Algebras}

To directly verify \eqref{dynamical_symmetry} 
(just using \eqref{def_Mb_minus})
is a very instructive, but complicated, calculation; 
in particular one finds that the modes of $\zeta$ (times $\eta$)
close under Poisson-brackets (on the constrained phase-space, i.e. assuming \eqref{divergencefree}),
\begin{equation} \label{zeta_Lie_algebra}
	\{ \eta \zeta_\alpha, \eta \zeta_{\alpha'} \} 
	= f_{\alpha\alpha'}^\epsilon \eta\zeta_\epsilon
\end{equation}
(whose simplest, $M=1$, example leads to the Virasoro-algebra).
Let me calculate the structure constants and identify 
the generators as special diffeomorphisms of $\Sigma_0$:

Using \eqref{int_condition_space}/\eqref{zeta_solved}
and \eqref{Greens_function}/\eqref{ON-basis}
(implying $G = \sum\limits_{\alpha=1}^\infty \frac{-1}{\mu_\alpha} 
Y_\alpha(\varphi)Y_\alpha(\tilde{\varphi})$)
one has (corresponding to a vectorfield whose divergence is $\nabla_a f^a = - Y_\alpha$)
\begin{equation} \label{def_eta_zeta}
	L_\alpha := \eta\zeta_\alpha 
	:= \int Y_\alpha \zeta\rho \thinspace d^M\varphi 
	= \frac{1}{\mu_\alpha} \int (\nabla^a Y_\alpha) \thinspace \vec{p} \cdot \partial_a \vec{x}
	= \int f_\alpha^a \thinspace \vec{p} \cdot \partial_a \vec{x};
\end{equation}
hence ( $\approx$ indicating the use of \eqref{divergencefree} , i.e. equal modulo volume-preserving diffeomorphisms )
\begin{eqnarray}
	\lefteqn{ \mu_\alpha \mu_{\alpha'} \{ \eta\zeta_\alpha, \eta\zeta_{\alpha'} \} } \nonumber \\
	&=& \left\{ \int \nabla^a Y_\alpha \vec{p} \cdot \partial_a \vec{x} \thinspace d^M\varphi,
		\int \nabla^{a'} Y_{\alpha'} \vec{p} \cdot \partial_{a'} \vec{x} \thinspace d^M\varphi' \right\} \nonumber \\
	&=& \int \left( \nabla^b Y_\alpha \nabla_b(\nabla^a Y_{\alpha'}) 
		- \nabla^b Y_{\alpha'} \nabla_b(\nabla^a Y_\alpha) \right)
		\vec{p} \cdot \partial_a \vec{x} \thinspace d^M\varphi \nonumber \\
	&\approx& -\int \left( \nabla^b Y_\alpha \nabla_a \nabla_b \nabla^a Y_{\alpha'}
		- (\alpha \leftrightarrow \alpha')
		\right) \eta \zeta \rho \thinspace d^M\varphi,
\end{eqnarray}
so that
\begin{equation} \label{structure_constants}
	f_{\alpha\alpha'}^\epsilon = \frac{\mu_{\alpha'} - \mu_\alpha}{2\mu_\alpha \mu_{\alpha'}} (\mu_\alpha + \mu_{\alpha'} - \mu_\epsilon) d_{\alpha\alpha'\epsilon}.
\end{equation}
For $M=1$ the combination of eigenvalues gives
$\frac{m^2 - n^2}{mn}$ which indeed 
(multiplying, in accordance with the conventional oscillator-expansions, the generators $L_m$ by m)
gives $(m-n)$.

Consequences of the dynamical symmetry, Lorentz-invariance in Matrix models, generalisations to the supersymmetric theories, and properties of the various algebras of local fields arising from $d_{\alpha\beta\gamma}$ and $e_{\alpha\beta\gamma}$ (cp. \eqref{e}) will be discussed in
forthcoming papers.

\subsubsection*{Acknowledgement}
I would like to thank M.Bordemann for innumerable discussions that, for many years, have always influenced my understanding.

\end{document}